\documentclass[final,3p,times,twocolumn]{elsarticle}

\usepackage{color}
\usepackage{epsfig}
\usepackage{amssymb}
\usepackage{amsmath}
\usepackage{dsfont}

\usepackage{graphicx,hyperref}
\usepackage{bm}
\usepackage{subfigure}

\newcommand{\beq}{\begin{equation}}
\newcommand{\eeq}{\end{equation}}
\newcommand{\bea}{\begin{eqnarray}}
\newcommand{\eea}{\end{eqnarray}}
\newcommand{\nn}{\nonumber \\}

\newcommand\eqn[1]{(\ref{#1})}      
\newcommand\Eqn[1]{Eq.~(\ref{#1})}  
\newcommand\Fig[1]{Fig.~\ref{#1}}  

\newcommand{\tr}{\hbox{tr}}

\newcommand{\cb}{{\bar c}}

\journal{Physics Letters B}

\begin{document}

\begin{frontmatter}



\title{Deconfinement transition in SU($N$) theories from perturbation theory}


\author[a]{U. Reinosa}
\author[b]{J. Serreau}
\author[c]{M. Tissier}
\author[d]{N. Wschebor}


\address[a]{Centre de Physique Th\'eorique, Ecole Polytechnique, CNRS, 91128 Palaiseau Cedex, France.}
\address[b]{Astro-Particule et Cosmologie (APC), CNRS UMR 7164, Universit\'e Paris 7 - Denis Diderot\\ 10, rue Alice Domon et L\'eonie Duquet, 75205 Paris Cedex 13, France.}
\address[c]{LPTMC, Laboratoire de Physique Th\'eorique de la Mati\`ere Condens\'ee, CNRS UMR 7600, \\Universit\'e Pierre et Marie Curie,  boite 121, 4 pl. Jussieu, 75252 Paris Cedex 05, France.}
\address[d]{Instituto de F\'{\i}sica, Facultad de Ingenier\'{\i}a, Universidad de la Rep\'ublica, \\J.H.y Reissig 565, 11000 Montevideo, Uruguay.}

\begin{abstract}

We consider a simple massive extension of the Landau-DeWitt gauge for SU($N$) Yang-Mills theory. We compute the corresponding one-loop effective potential for a temporal background gluon field at finite temperature. At this order the background field is simply related to the Polyakov loop, the order parameter of the deconfinement transition. Our perturbative calculation correctly describes a quark confining phase at low temperature and a phase transition of second order for $N=2$ and weakly first order for $N=3$. Our estimates for the transition temperatures are in qualitative agreement with values from lattice simulations or from other continuum approaches. Finally, we discuss the effective gluon mass parameter in relation to the Gribov ambiguities of the Landau-DeWitt gauge.

\end{abstract}

\begin{keyword} Yang-Mills theory \sep finite temperature QFT \sep deconfinement transition \sep Gribov ambiguities


\end{keyword}

\end{frontmatter}



\section{Introduction}\label{sec:introduction}

At finite temperature, SU($N$) Yang-Mills theories present a phase transition associated with the spontaneous breaking of the $Z_N$ center symmetry \cite{Svetitsky:1985ye} with a transition temperature roughly of the order of the intrinsic scale of the theory \cite{Fingberg:1992ju}. The Polyakov loop provides a relevant order parameter for this transition which can be related to the free energy of an isolated static quark \cite{Polyakov:1978vu}. At high temperature, where center symmetry is broken, the Polyakov-loop admits a non zero value corresponding to a finite free energy and thus to a quark deconfined phase. As the temperature is lowered, center symmetry is eventually restored, the Polyakov loop vanishes and the free energy becomes infinite signaling the presence of a quark confining phase. It is widely accepted that this phenomenon, which has its origins in the infrared (IR) dynamics of the theory, is intrinsically nonperturbative. This is based on the failure of standard perturbative techniques for energy scales of the order of the transition temperature. Existing calculations involve either direct lattice simulations \cite{Svetitsky:1985ye,Kaczmarek:2002mc,Lucini:2005vg,Greensite:2012dy,Smith:2013msa} or nonperturbative semi-analytical techniques such as the functional renormalization group (FRG) \cite{Marhauser:2008fz,Braun:2007bx,Braun:2010cy,Fister:2013bh}, Schwinger-Dyson equations \cite{Epple:2006hv,Alkofer:2006gz,Fischer:2009wc,Fischer:2009gk}, the Hamiltonian approach of \cite{Reinhardt:2012qe}, or a two-particle-irreducible inspired approach \cite{Fukushima:2012qa,Fister:2013bh}. Another interesting approach is that based on the (refined) Gribov-Zwanziger action \cite{Zwanziger:1989mf,Dudal:2008sp}. To our knowledge, it has been used to study the behavior of thermodynamic quantities across the transition \cite{Zwanziger:2004np,Zwanziger:2006sc,Fukushima:2013xsa,Canfora:2013kma} but, so far, not to investigate the issue of center-symmetry restoration at low temperatures.

In the present Letter, we  show that the main aspects of the $Z_N$ transition can be described by means of a slightly modified perturbative expansion. Our approach is based on a series of recent developments concerning the description of the infrared sector of gauge-fixed Yang-Mills theories \cite{Tissier_10,Tissier_11,Serreau:2012cg,Pelaez:2013cpa,Serreau:2013ila,Reinosa:2013twa,Pelaez:2014mxa}. Lattice calculations of gluon and ghost-antighost correlators in the Landau gauge, both in the vacuum and at finite temperature, show that at low momenta, the gluon behaves as a massive field whereas the ghost and antighost remain massless. This has motivated two of us to study a simple massive extension of the Faddeev-Popov Landau gauge Lagrangian \cite{Tissier_10,Tissier_11}, which is a special case of the Curci-Ferrari model \cite{Curci76} also known as the Fradkin-Tyutin model \cite{Fradkin:1970ib}. A one-loop calculation reproduces the lattice data for the vacuum two-point correlators  with very good accuracy all the way from high momenta down to the deep infrared regime. The effective gluon mass also provides an infrared regulator which allows for infrared safe renormalization group trajectories with no Landau pole. A similar one-loop calculation of the vacuum three-point correlators in the Landau gauge has been performed recently \cite{Pelaez:2013cpa} and gives equally good results, although the lattice data are more noisy. This demonstrates that the dominant infrared physics (in the vacuum and in the Landau gauge) is efficiently captured by a simple effective gluon mass. 

A qualitative but instructive analogy can be made with the physics of the Kosterlitz-Thouless transition in the $XY$ model in statistical physics \cite{Kosterlitz:1973xp,Jose:1977gm}. In terms of usual spin waves, the physics of the transition is intrinsically nonperturbative: perturbation theory shows no sign of the transition at all orders. Alternatively, taking explicitly into account the vortex excitations of the $XY$ model leads to a valid description of the transition by perturbative means. The validity of a perturbative description crucially depends on having a physically sound reference action. 

The analogy with (gauge-fixed) Yang-Mills theories lies in the fact that the Faddeev-Popov action completely ignores the existence of Gribov ambiguities and is, at best, a valid description at asymptotically high energies, where the latter play a negligible role. In this formulation, the description of low energy phenomena is a problem of notorious nonperturbative nature. Instead, the massive extension of the Faddeev-Popov Lagrangian seems to provide a more efficient starting point for the (perturbative) description of low energy phenomena. This may be due to the fact that a complete gauge fixing procedure, which consistently deals with the Gribov issue, is likely to explicitly break the BRST symmetry and thus to induce (effective) BRST breaking terms. The simplest such term consistent with the requirement of renormallizability and locality is a gluon mass term.

Such qualitative ideas can be given more solid theoretical foundations. Explicit examples in the Landau gauge are the (refined) Gribov-Zwanziger approach \cite{Zwanziger:1989mf,Dudal:2008sp}, where one restricts the path integral to the first Gribov region, or the quantization procedure of \cite{Serreau:2012cg}, where one performs a particular average over Gribov copies. Both approaches induce a soft breaking of the BRST symmetry.\footnote{The minimal Landau gauge on the lattice, where one selects a particular Gribov copy, provides another example, although less explicit since it has no known continuum formulation.} In the latter case, the resulting gauge-fixed Yang-Mills action has been shown to be perturbatively equivalent to the Curci-Ferrari-Fradkin-Tyutin (CFFT) model for what concerns the calculation of gluon and ghost-antighost correlators. In this context, the (bare) gluon mass appears as a gauge-fixing parameter which lifts the degeneracy between the Gribov copies. The one-loop calculations of Yang-Mills correlators mentioned above can be viewed as genuine applications of perturbation theory in this class of gauges.

More recently, we have investigated to what extent the above results hold at finite temperature \cite{Reinosa:2013twa} by comparing to various lattice results \cite{Cucchieri11,Aouane:2011fv,Maas:2011ez,Silva:2013maa}.  We have performed a one-loop calculation of the gluon (electric and magnetic) and ghost-antighost propagators, fitting the mass and coupling parameters against the lattice data for each temperature. We found a quantitative agreement for the ghost and magnetic gluon propagators, demonstrating that a simple mass term is enough to capture the corresponding physics all the way to deep infrared momenta. As for the electric gluon propagator, the results are quantitatively good except near the transition temperature, where the agreement is only qualitative. It must be mentioned though that lattice data are pretty noisy in this region \cite{Cucchieri11,Cucchieri12,Mendes:2014gva}. These results are similar to those from other continuum approaches \cite{Fister:2011uw}. 

However, an important physical ingredient is missing in these continuum calculations, namely the explicit $Z_N$ symmetry and the existence of an associated order parameter. In fact, this symmetry is explicitly broken in the Landau gauge and, although this should not affect physical observables in an exact calculation, it questions the ability of approximate approaches to capture the physics of the transition. Besides, this may also explain the poor convergence of lattice results for, say, the Debye mass near the transition temperature \cite{Mendes:2014gva}.

It is, therefore, desirable to devise an approach where the $Z_N$ symmetry can be explicitly accounted for both on the lattice and in (approximate) continuum calculations. Background field gauge (BFG) methods provide an efficient way to go since the center symmetry is explicit at the quantum level and easy to maintain in approximation schemes. In Refs. \cite{Braun:2007bx,Braun:2010cy,Fister:2013bh}, the authors have studied the $Z_N$ deconfinement phase transition by means of FRG techniques in the BFG formalism in the Landau-DeWitt (LDW) gauge, the background field generalization of the Landau gauge. This approach correctly predicts the order of the phase transition for $N=2$ and $N=3$ and gives reasonable transition temperatures.

In the present Letter, we follow this path and investigate whether the perturbative approach described above, based on a simple gluon mass, is able to capture the physics of the deconfinement transition in the BFG formalism. More precisely, we implement a simple massive extension of the LDW gauge. This is in fact equivalent to the CFFT model in the presence of a source associated with a composite operator $s\bar s A_\mu$, where $s$ and $\bar s$ denote appropriate BRST and antiBRST variations \cite{Tissier:2008nw,Binosi:2013cea}. The quantum action in the presence of this source satisfies a new Ward identity \cite{Tissier:2008nw} which actually corresponds to the background field gauge symmetry in the BFG formulation \cite{Binosi:2013cea}. As we shall describe below, this symmetry is what allows one to keep track of the underlying center symmetry. In the BFG formalism, observables such as the thermodynamic pressure or the Polyakov loop can be computed unambiguously by evaluating their functional expression at the minimum of a particular background field potential. 

We compute this potential at one-loop order in the massive version of the LDW gauge and follow its minimum as a function of the temperature. We probe the $Z_N$ symmetry by computing the Polyakov loop evaluated at this minimum. At the order of approximation considered here, it coincides with the mean field expression considered in previous FRG approaches. Although, this identification does not hold beyond leading order, the mean field Polyakov loop is expected to provide an equally valid order parameter \cite{Marhauser:2008fz,Braun:2007bx}.\footnote{This has been proved for the SU($2$) theory in the Polyakov gauge in \cite{Marhauser:2008fz}. However, to our knowledge, no proof exists in the LDW gauge. Alternative order parameters have also been proposed; see, e.g., \cite{Gattringer:2006ci}.}

Thanks to the gluon mass, the gluon loop contribution is suppressed at low temperatures and the potential is dominated by the massless ghost loop. This leads to a center symmetric minimum, both in SU($2$) and SU($3$), hence to a confined phase, in accordance with the results of \cite{Braun:2007bx}. At higher temperatures, a non-trivial, center symmetry breaking minimum develops. We find a second order transition in SU($2$) and a first order one in SU($3$), as expected, with transition temperatures controlled by the gluon mass. We estimate the latter from fits of lattice data for the vacuum gluon propagator in the Landau gauge. Our one-loop calculation essentially reproduces the results of the FRG approach of Ref. \cite{Braun:2007bx}.  We emphasize that the present calculation can be systematically improved by computing higher order contributions.

Finally, we discuss the relation of the gluon mass with the issue of Gribov ambiguities in the LDW gauge. The quantization procedure proposed in \cite{Serreau:2012cg} in the Landau gauge can be generalized to the LDW gauge in a straightforward way. The resulting gauge-fixed Yang-Mills action is perturbatively equivalent to the massive model discussed in the core of the paper as far as gluon and ghost-antighost correlators are concerned. As in the Landau gauge, the gluon mass term appears as a tool to control the Gribov problem and completely fixes the gauge.

\section{General set-up}

\subsection{Massive background field action and $Z_N$ symmetry}

We consider a SU($N$) Yang-Mills theory in $d$ dimensions with Euclidean time $\tau\in[0,\beta]$, where $\beta=1/T$ is the inverse temperature. The classical action reads 
\beq 
\label{eq:YM}
S_{\rm  YM}[A]=\frac{1}{4}\int_x F^a_{\mu\nu}F_{\mu\nu}^a\,, 
\eeq 
where $F_{\mu\nu}^a=\partial_\mu A_\nu^a-\partial_\nu
A_\mu^a +gf^{abc}A_\mu^bA_\nu^c$, with $g$ the coupling constant, and $\int_x=\int_0^\beta d\tau\int d^{d-1} x$. 
Our convention for the SU($N$) generators is $ \tr \,(t^a t^b)=\frac 12 \delta^{ab}$ and $[t^a,t^b]=if^{abc}t^c$. We define matrix fields as $\varphi=\varphi^a t^a$.

We quantize the theory using the background field method \cite{Weinberg:1996kr}, which introduces an {\it a priori} arbitrary background field configuration $\bar A_\mu$ through a modified gauge-fixing condition. In terms of $a_\mu=A_\mu-\bar A_\mu$, the LDW gauge condition reads
\beq
\label{eq:LdW}
 \bar D_\mu a_\mu^a=0,
\eeq
where $\bar D_\mu^{ab}=\partial_\mu \delta^{ab}+g f^{acb}\bar A_\mu^c$. Here and in the following, we use the notation $\bar D_\mu \varphi^a=\bar D_\mu^{ab} \varphi^b$. The standard Faddeev-Popov gauge-fixed action reads \cite{Binosi:2013cea}
\beq
\label{eq_gf}
 S^{\!\rm gf}_{\bar A}=\!\int_x\left\{{1\over4}F_{\mu\nu}^aF_{\mu\nu}^a+\bar D_\mu\bar c^aD_\mu c^a+ih^a\bar D_\mu a_\mu^a\right\}  ,
\eeq
with a (real) Nakanishi-Lautrup field $h$ and ghosts and antighost fields $c$ and $\bar c$. Note that, in terms of the field $a_\mu$, one has $F_{\mu\nu}^a=\bar F_{\mu\nu}^a+\bar D_\mu a_\nu^a-\bar D_\nu a_\mu^a+g f^{abc}a_\mu^b a_\nu^c$, with $\bar F_{\mu\nu}^a$ the background field strength tensor, and $D_\mu^{ab}=\bar D_\mu^{ab}+g f^{acb}a_\mu^c$.

The action \eqn{eq_gf} has the obvious symmetry
\beq
\label{eq:bckgsymbare}
 S^{\!\rm gf}_{\bar A}[\varphi]=S^{\!\rm gf}_{\bar A^U}[U\varphi U^{-1}]
\eeq
where $\varphi=(a,c,\bar c,h)$, $U$ is a local SU($N$) matrix and 
\begin{equation}
\label{eq:bckgtrans}
  \bar A_\mu^U=U\bar A_\mu U^{-1}+\frac i {g}U\partial_\mu U^{-1}.
\end{equation}
Here, we consider a simple massive extension of \eqn{eq_gf} which preserves the symmetry:
\beq
\label{eq_CF}
 S_{\bar A}=\!\int_x\left\{{1\over4}F_{\mu\nu}^aF_{\mu\nu}^a+{m^2\over2}a_\mu^aa_\mu^a+\bar D_\mu\bar c^aD_\mu c^a+ih^a\bar D_\mu a_\mu^a\right\}\!.
\eeq
It is easy to check that the linear symmetry \eqn{eq:bckgsymbare} implies the corresponding one at the level of the (quantum) effective action $\Gamma$ \cite{Weinberg:1996kr}:
\beq
\label{eq_ginv}
 \Gamma_{\bar A}[\varphi]=\Gamma_{\bar A^U}[U\varphi U^{-1}],
\eeq
where the collection of fields in $\varphi$ are now to be understood as average values in the presence of sources.\footnote{In particular $h$ is now purely imaginary.} The symmetry (\ref{eq:bckgsymbare}) is trivially preserved in perturbation theory. In particular, it encodes the relevant center symmetry governing the deconfinement transition at finite 
temperature.

\subsection{Observables and the effective potential}

Before embarking in actual calculations, we recall how gauge-invariant observables, such as the thermodynamic pressure or the Polyakov loop to be discussed below, can be conveniently obtained by choosing a particular background field $\bar A=\bar A_{\rm min}$ which minimizes a specific background field potential defined below. For the sake of the argument, let us consider the case of vanishing sources for the fields $c$, $\cb$ and $h$. We thus consider the effective action $\Gamma_{\bar A}[a]$ for the field $a$ and a given background $\bar A$. 

A physical (gauge-invariant) observable $\cal O$ can be obtained as the value of a given functional ${\cal O}_{\bar A}[a]$ evaluated for $a$ at the absolute minimum of the functional $\Gamma_{\bar A}[a]$:
\beq
\label{eq:eom}
 {\cal O}={\cal O}_{\bar A}[a^{\rm min}_{\bar A}],
\eeq
where $a^{\rm min}_{\bar A}$ satisfies
\beq
\label{eq_absmin}
 \Gamma_{\bar A}[a^{\rm min}_{\bar A}]\le\Gamma_{\bar A}[a]\quad\forall a.
\eeq
Note that, by definition, physical observables are independent of the gauge-fixing condition and thus of $\bar A$. In particular, the effective action evaluated at its absolute minimum is directly related to the partition function $Z$ and is thus gauge invariant: 
\beq
\label{eq_partition}
 - \ln Z= \Gamma_{\bar A}[a^{\rm min}_{\bar A}] \quad\forall\bar A.
\eeq

Taking advantage of the background field independence of observables, it is judicious to choose $\bar A=\bar A_{\rm min}$, defined as $a^{\rm min}_{\bar A_{\rm min}}=0$. We then have 
\beq
 {\cal O}={\cal O}_{\bar A_{\rm min}}[0].
\eeq
Indeed, consider the following functional of the background field\footnote{Notice that $\tilde \Gamma[\bar A]$ is {\em not} a Legendre transform.}
\beq
\label{eq_functilde}
 \tilde\Gamma[\bar A]=\Gamma_{\bar A}[0],
\eeq
which is invariant under the gauge transformation \eqn{eq:bckgtrans}: $ \tilde\Gamma[\bar A]= \tilde\Gamma[\bar A^U]$, as follows from \Eqn{eq_ginv}.
One easily proves that $\bar A_{\rm min}$, defined above, is an absolute minimum of $\tilde\Gamma[\bar A]$:
\beq 
 \tilde\Gamma[\bar A_{\rm min}]=\Gamma_{\bar A_{\rm min}}[a^{\rm min}_{\bar A_{\rm min}}]=\Gamma_{\bar A}[a^{\rm min}_{\bar A}]\le\Gamma_{\bar A}[0]=\tilde\Gamma[\bar A]\quad\forall \bar A.
\eeq
Here, the first equality just states the definitions of $\tilde \Gamma$ and of $\bar A_{\rm min}$, the second one follows from the gauge invariance of the partition function, \Eqn{eq_partition}, and the inequality is just \Eqn{eq_absmin} taken for $a=0$.

For our present purposes, it is sufficient to consider homogeneous background fields in the temporal direction. We thus consider the background field potential 
\beq
\label{eq:poteff}
 V^{\rm SU(N)}(T,\beta g\bar A)=\frac{\tilde\Gamma[\bar A]}{\beta \Omega},
\eeq
where $\Omega$ is the spatial volume. The thermodynamic pressure is then simply given by 
\beq
\label{eq:pressure}
 p=-V^{\rm SU(N)}(T,\beta g\bar A_{\rm min}),
\eeq
whereas the Polyakov loop is obtained as
\beq
\label{eq:ploop}
 \ell=\frac{1}{N}\tr \left< P\exp\left\{ig\int_0^\beta d\tau \left[\bar A_0^{\rm min}+a_0(\tau,{\bf x})\right]\right\}\right>_{\rm min}
\eeq
where $P$ denotes the path ordering (operators are ordered from left to right according to the decreasing value of their time argument) and where the subscript 'min' means that the average must be evaluated for $\langle a\rangle=a^{\rm min}_{\bar A_{\rm min}}=0$. 

\section{The background field potential at one loop}

We consider a constant temporal background field $\bar A_\mu(x)=\bar A_0\delta_{\mu0}$. The Hermitian matrix $\bar A_0$ can be diagonalized by means of a global SU($N$) rotation and we can thus, with no loss of generality, consider the case where $\bar A_0$ belongs to the Cartan subalgebra of the gauge group: $A_0\to A_0^at_C^a$ with $[t_C^a,t_C^b]=0$.

At lowest non-trivial (one-loop) order in a perturbative expansion, the potential \eqn{eq:poteff} reads
\beq
\label{eq:potoneloop}
 V^{\rm SU(N)}(T,r_a)=\frac{1}{\beta \Omega}\left\{\frac{1}{2}{\rm Tr}\,{\rm Ln}\,\Delta^{-1}_{a,h}-{\rm Tr}\,{\rm Ln}\,\Delta^{-1}_{c,\bar c}\right\},
\eeq
where $r_a=\beta g \bar A_0^a$ and where $\Delta^{-1}_{a,h}$ and $\Delta^{-1}_{c,\bar c}$ denote the tree-level inverse propagator in the $(a,h)$ and $(c,\bar c)$ sectors respectively. At the same order of approximation, the Polyakov loop \eqn{eq:ploop} assumes its tree-level expression
\beq
\label{eq:poltreelevel}
 \ell=\frac{1}{N}\tr\left[\exp\left(ir_at^a\right)\right].
\eeq
where it is understood that the right-hand side must be evaluated at the absolute minimum of the potential \eqn{eq:potoneloop}. In the following, we compute explicitly the potential and the Polyakov loop in the cases $N=2$ and $N=3$.

\subsection{Second order transition for $N=2$}

The Cartan algebra has one single direction, say $a=3$ and we note $r_3=r$. As discussed in the Appendix, the background field potential is $2\pi$-periodic in $r$ as a result of the $Z_2$ center symmetry implies. Together with charge-conjugation invariance, this implies
\beq
\label{eq:symSU2}
 V^{\rm SU(2)}(T,\pi +r)=V^{\rm SU(2)}(T,\pi-r).
\eeq
It is thus sufficient to study the potential in the interval $r\in[0,\pi]$ and we see that both $r=0$ and $r=\pi$ are always extrema. The tree-level Polyakov loop \eqn{eq:poltreelevel} reads
\beq
 \ell=\cos(r/2).
\eeq
One concludes that $r\in [0,\pi[$ and $r=\pi$ correspond respectively to $Z_2$-breaking and $Z_2$-symmetric solutions.

The calculation of the one-loop expression \eqn{eq:potoneloop} is straightforward. Introducing the notations $n^a=r^a/r$ and  $R_\mu=rT\delta_{\mu0}$, the covariant derivative $\bar D$ reads, in momentum space,\footnote{Our sign convention for Fourier transforms is $f(x)=\int dk e^{-ikx}\tilde f(k)$ and similarly for discrete Fourier series in the compact Euclidean time direction.}
\beq
\label{eq:covder}
 i\bar D_\mu^{ab}\rightarrow Q_\mu {\cal P}_0^{ab}+(Q+R)_\mu {\cal P}_+^{ab}+(Q-R)_\mu {\cal P}_-^{ab},
\eeq
where $Q_\mu=(\omega_n,{\bf q})$, with $\omega_n=2\pi n T$ the Matsubara frequencies and where we introduced the set of orthogonal Hermitian projectors
\beq
 {\cal P}_0^{ab}=n^a n^b\,, \quad {\cal P}_\pm^{ab}=\frac{1}{2}\left(\delta^{ab}-n^an^b\pm i\varepsilon^{acb}n^c\right),
\eeq
which satisfy ${\cal P}_\eta{\cal P}_{\eta'}=\delta_{\eta\eta'}{\cal P}_\eta$ and $\sum_{\eta=0,\pm}{\cal P}_\eta=\mathds{1}$.

\begin{figure}[t!]  
\begin{center}
\hspace*{-.5cm}\epsfig{file=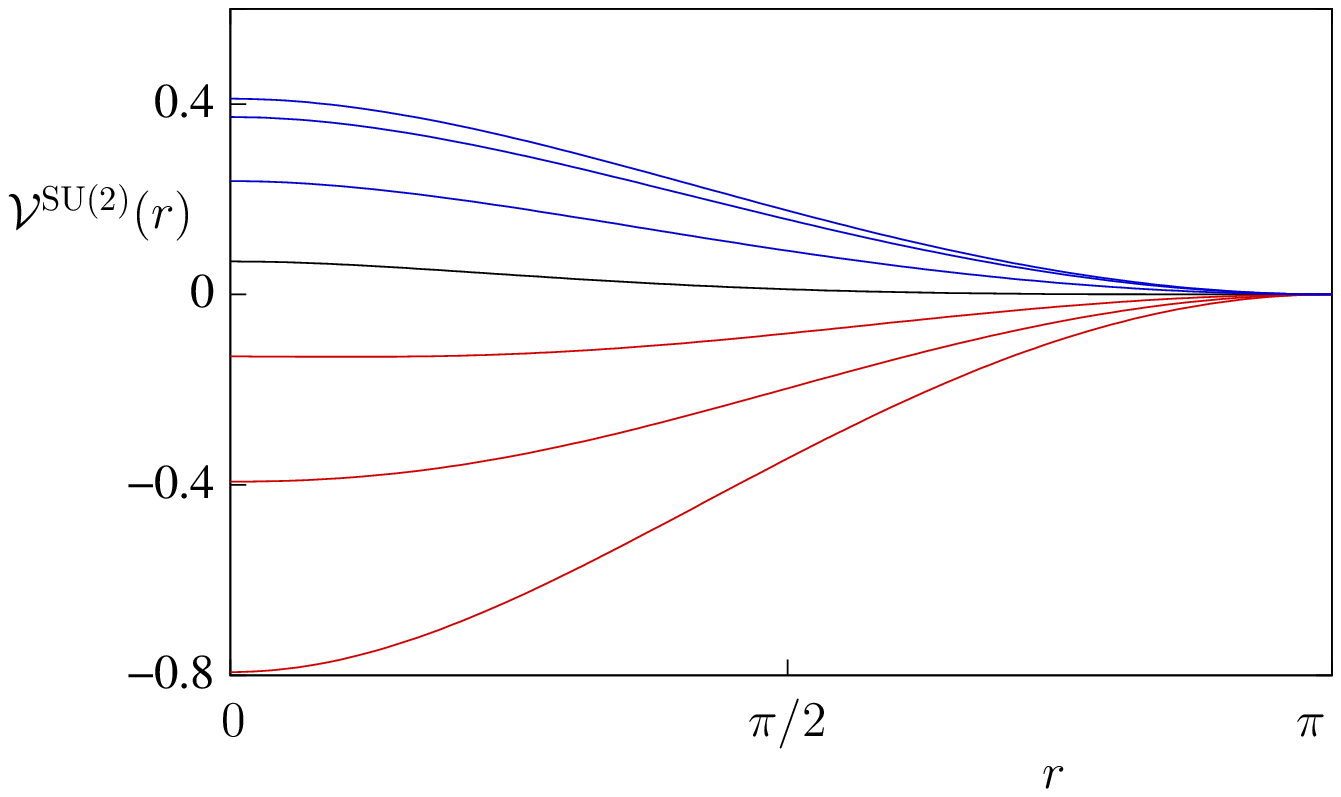,width=6.2cm}\\
\vspace{0.3cm}
\hspace*{-.5cm}\epsfig{file=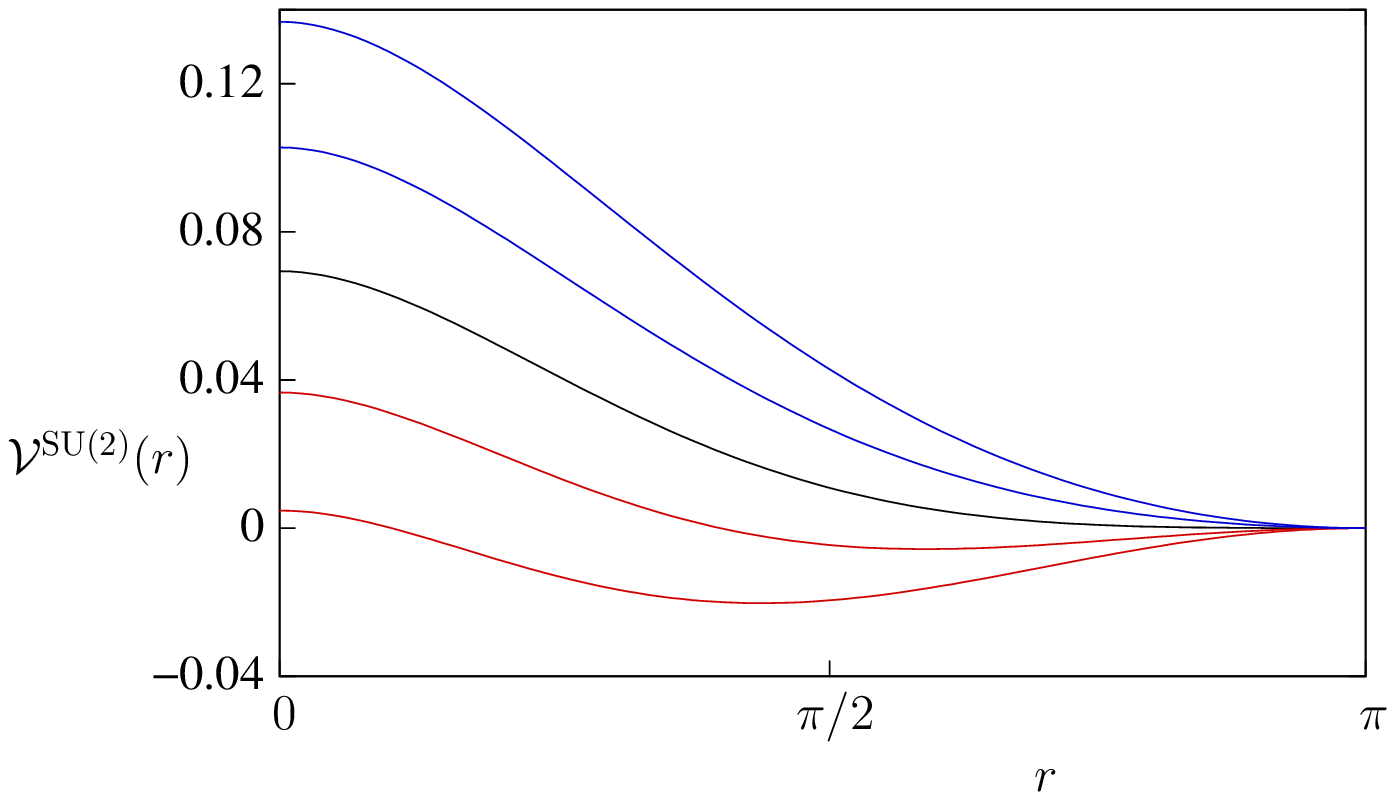,width=6.2cm}
 \caption{\label{fig:SU2pot} The background field potential \eqn{eq:sub} in $d=4$, normalized to its value at $r=\pi$, for decreasing values of $\tilde m$ (increasing temperatures) from top to bottom. The bottom figure is a close-up view around $T_{\rm c}$. (color online: $T=T_{\rm c}$ (black), $T<T_{\rm c}$ (blue), $T>T_{\rm c}$ (red))}
\end{center}
\end{figure}

One can consider each orthogonal color sector independently. The covariant derivative \eqn{eq:covder} is block-diagonal with one $1\times1$ block along the Cartan direction (${\cal P}_0$) and one $2\times2$ block in the transverse directions (${\cal P}_\pm$). Since in the action \eqn{eq_CF} the background field only appears in the covariant derivative \eqn{eq:covder}, we see that in each color sector the calculation of the inverse propagators $\Delta^{-1}_{a,h}$ and $\Delta^{-1}_{c,\bar c}$ and thus of the one-loop expression \eqn{eq:potoneloop} is formally equivalent to considering the case of an imaginary chemical potential $\mu=irT$ with corresponding U($1$) charges $\eta=0,\pm$. 
Introducing the notation
\beq
\label{eq:Msum}
 T^d F_{\tilde m}(r)\,\,\hat{=}\,\,T\sum_{n\in\mathbb{Z}}\int_{\bf q}\ln[(Q+ R)^2+m^2],
\eeq
where $\tilde m=m/T$, $\int_{\bf q}=\int d^{d-1}q/(2\pi)^{d-1}$, and $\hat{=}$ means that we subtract a (divergent) term independent of $T$ and $r$, we have
\begin{align}
&\frac{{\rm Tr}\,{\rm Ln}\,\Delta^{-1}_{a,h}}{\beta \Omega}\,\,\hat{=}\,\,T^d\!\!\sum_{\eta=0,\pm}\!\left[(d-1)F_{\tilde m}(\eta r)+F_0(\eta r)\right]\!,\\
&\frac{{\rm Tr}\,{\rm Ln}\,\Delta^{-1}_{c,\bar c}}{\beta \Omega}\,\,\hat{=}\,\,T^d\!\!\sum_{\eta=0,\pm}F_0(\eta r).
\end{align}
The Matsubara sum in \eqn{eq:Msum} is easily evaluated as
\beq
\label{eq:F}
 F_{\tilde m}(r)=\int_{\bf q}\ln\left(1+e^{-2\tilde \varepsilon_q}-2e^{-\tilde \varepsilon_q}\cos r\right)\,,
\eeq
where $\tilde \varepsilon_q=\sqrt{{\bf q}^2+\tilde m^2}$. Introducing the function
\beq
\label{eq:W}
 {\cal W}_{\tilde m}(r)=\frac{1}{2}\Big[(d-1)F_{\tilde m}(r)-F_{0}(r)\Big],
\eeq
our final expression for  the dimensionless potential ${\cal V}^{\rm SU(2)}(T,r)\equiv V^{\rm SU(2)}(T,r)/T^d$ is
\beq
\label{eq:sub}
{\cal V}^{\rm SU(2)}(T,r)\,\,\hat{=}\,\,2{\cal W}_{\tilde m}(r)+{\cal W}_{\tilde m}(0).
\eeq
The two terms on the right-hand side come from the transverse and longitudinal color modes respectively. The potential \eqn{eq:sub} is plotted in \Fig{fig:SU2pot}.

In the high temperature limit $\tilde m\ll1$, one has ${\cal W}_{\tilde m}(r)\approx(d-2)F_0(r)/2$ and we recover the standard Weiss potential \cite{Weiss:1980rj}
\beq
\label{eq:highT}
 {\cal V}^{\rm SU(2)}(T,r)\approx \frac{d-2}{2}\left[2F_0(r)+F_0(0)\right],
\eeq
with 
\begin{align}
 F_0( r)=-2\frac{\Omega_{d-1}}{(2\pi)^{d-1}}\Gamma(d-1)\sum_{n\ge1}\frac{\cos nr}{n^d},
\end{align}
where $\Omega_D=2\pi^{D/2}/\Gamma(D/2)$. In particular, in $d=4$, this reads
\beq
 F_0^{d=4}( r)=\frac{(r-\pi)^4}{24\pi^2}-\frac{(r-\pi)^2}{12}+\frac{7\pi^2}{360}\quad{\rm for}\quad r\in[0,2\pi].
\eeq
The function \eqn{eq:highT} presents a $Z_2$-breaking minimum at $r=0$. The extremum at $r=\pi$ is a maximum.

At low temperatures, the contribution from massive gluons in \Eqn{eq:W} is exponentially suppressed, leaving only the ghost contribution: ${\cal W}_{\tilde m}(r)\approx-F_0(r)/2$ and we get the inverted Weiss potential\footnote{This is in line with the general arguments developed in \cite{Braun:2007bx,Fister:2013bh}.}
\beq 
 {\cal V}^{\rm SU(2)}(T, r)\approx -\frac{1}{2}\left[2F_0(r)+F_0(0)\right].
\eeq
This exhibits a $Z_2$-symmetric minimum at $ r=\pi$.
As the temperature increases, the potential flattens around the minimum $r=\pi$ and a new minimum eventually develops, as shown in Fig. \ref{fig:SU2pot}. There is a continuous transition at a critical temperature $T_c$ defined by
\beq
 \left.\frac{d^2}{d r^2}{\cal V}^{\rm SU(2)}(T_c, r)\,\right|_{ r=\pi}=0,
\eeq
with $\tilde m_c=m/T_c$. A simple calculation yields\footnote{Note that \Eqn{eq:F} involves (using an integration by parts) $f(x)={\rm Re} \,n_{\rm BE}(x+ir)$, with $n_{\rm BE}(x)=(e^x-1)^{-1}$ the Bose-Einstein distribution function. As emphasized before, $r$ appears as an imaginary chemical potential. In the high temperature phase, $r=0$ and this reduces to the standard Bose-Einstein function: $f(x)=n_{\rm BE}(x)$. Instead, in the low temperature, confining phase, $r=\pi$, and the transverse gluons and ghosts obey an abnormal Fermi-Dirac statistics: $f(x)=-n_{\rm FD}(x)$.}
\beq
 (d-1)\!\!\int_{\bf q}n_{\rm FD}'(\tilde \varepsilon_q)\,\Big|_{\tilde m_c}=\int_{\bf q}n_{\rm FD}'(q)\equiv-\frac{\Omega_{d-1}}{(2\pi)^{d-1}}h_d
,
\eeq
where $n_{\rm FD}(x)=(e^x+1)^{-1}$ and the prime is a derivative. The evaluation of the massless integral $h_d$ is straightforward and gives, for instance, $h_2=1/2$, $h_3=\ln 2$, $h_4=\pi^2/6$ and, more generally,
\beq
 h_d=\left(1-\frac{1}{2^{d-3}}\right)\Gamma(d-1)\zeta(d-2)\quad {\rm for}\quad d\ge3.
\eeq
We find\footnote{Interestingly, the above analysis implies $\tilde m_c=0$ in $d=2$: There is no phase transition and the system is always in the symmetric (confining) phase. We mention, however, that the massive model considered here is problematic when $d=2$ \cite{Tissier_10,Tissier_11}.} $\tilde m_c\simeq 2.97$ in $d=4$ and $\tilde m_c\simeq 1.72$ in $d=3$. 

In principle, estimates for the gluon mass can be inferred from the value of an independent observable, such as the glueball mass, or from gauge-fixed lattice data for, say, the Yang-Mills propagators. In full generality, this mass parameter, e.g., when viewed as effectively arising from a proper account of Gribov copies, may depend on both the temperature and the background field. No lattice calculations exist so far in the LDW gauge used in the present work, but we can use estimates from the $T=0$ lattice data in the Landau gauge \cite{Cucchieri11}, in the spirit of Refs. \cite{Tissier_10,Tissier_11}. This makes sense because the minimum of the background field potential scales as $T$ and vanishes in the vacuum: The LDW gauge thus reduces to the Landau gauge. 

In a previous work \cite{Reinosa:2013twa}, we could study the temperature dependence of the gluon mass by fitting the gluon and ghost correlators at finite temperature in the Landau gauge to existing lattice data. This showed almost no dependence in the range of temperature considered. Although we have no guaranty that this remains true in the present case, it indicates that an estimate at $T=0$ is probably not completely off.

The fits performed in Refs. \cite{Tissier_10,Tissier_11} used one-loop propagators which is beyond the order of approximation of the present calculation. To be consistent, we have performed fits using the tree-level vacum propagators. In $d=4$ we obtain $m=710$~MeV, which gives $T_c\simeq 238$~MeV. A direct comparison to values of $T_c$ from lattice studies or from other continuum approaches is difficult because of the issue of properly setting the scale. A typical lattice value \cite{Lucini:2012gg} is $T_c=295$~MeV and the most recent value from FRG studies \cite{Fister:2013bh} is $T_c=230$~MeV. Although, as mentioned above, our above estimate is only qualitative, our one-loop calculation falls in the right ballpark.

\subsection{First order transition for $N=3$}

We now consider the case $N=3$. The Cartan subalgebra has two directions, $a=3,8$ in the usual Gell-Mann representation. As discussed in Appendix \ref{sym}, the $Z_3$-center symmetry puts no constraint in the $r_3$-direction. Standard (periodic) gauge transformations in that direction leads to a $4\pi$-periodicity in $r_3$. In contrast, the $Z_3$-symmetry gives a shorter periodicity of $4\pi/\!\sqrt{3}$ in the $r_8$ direction. We thus have 
\begin{align}
 V^{\rm SU(3)}(T,r_3,r_8)&=V^{\rm SU(3)}(T,r_3,r_8+4\pi/\!\sqrt{3})\nn
 &=V^{\rm SU(3)}(T,r_3+4\pi,r_8).
\end{align}
As described in the Appendix, charge conjugation invariance and the global color symmetry further imply 
\begin{align}
& V^{\rm SU(3)}(T,r_3,r_8)=V^{\rm SU(3)}(T,-r_3,-r_8)\nn
 &=V^{\rm SU(3)}(T,r_3',r_8')=V^{\rm SU(3)}(T,-r_3,r_8),
\end{align}
where $(r_3',r_8')$ is obtained from $(r_3,r_8)$ by a rotation of angle $\pm\pi/3$. It is thus sufficient to study an equilateral triangle of side $4\pi/\!\sqrt{3}$ in the $(r_3,r_8)$ plane, with vertices at $(0,0)$ and $(2\pi,\pm 2\pi/\!\sqrt{3})$. Furthermore, the above symmetries imply that the potential is invariant under rotations which leave this equilateral triangle invariant, such that only a third of it is actually relevant. Finally, this implies that the vertices of the triangle and its center, located at $(4\pi/3,0)$, are always extrema of the background field potential. 

The expression of the tree-level Polyakov loop \eqn{eq:poltreelevel} reads
\beq
 \ell=\frac{1}{3}\left[e^{-i\frac{r_8}{\sqrt{3}}}+2e^{i\frac{r_8}{2\sqrt{3}}}\cos(r_3/2)\right],
\eeq
from which one sees that the vertices of the triangle correspond to $Z_3$-breaking solutions, with $\ell=\exp(2ni\pi/3), n\in \mathds{Z}$, whereas its center corresponds to the $Z_3$-symmetric solution, with $\ell=0$.
We shall see below that the absolute minimum of the one-loop background field potential lies on the $r_8=0$ axis, up to rotations of $\pm\pi/3$ as mentioned above, where the tree-level Polyakov loop is real:\footnote{This is consistent with the fact that the Polyakov loop is proportional to $\exp(-\beta F)$, where $F$ is the free energy of a static quark.} 
\beq
 \ell=\frac{1+2\cos(r_3/2)}{3}.
\eeq

The calculation of the background field potential follows similar lines as in the SU($2$) case. The covariant derivative $i\bar D^{ab}$ is block-diagonal with two $1\times1$ blocks corresponding to the Cartan directions and three $2\times2$ blocks with similar structures as in the SU($2$) case, corresponding to the three SU($2$) subgroups of SU($3$). The result can thus be written in terms of the function \eqn{eq:W}. Ignoring, again, a constant contribution,  
we obtain, for the dimensionless potential ${\cal V}^{\rm SU(3)}\equiv{V^{\rm SU(3)}/T^d}$
\begin{align}
\label{eq:potsu3}
  {\cal V}^{\rm SU(3)}\,\,\hat{=}\,\,2\left[{\cal W}_{\tilde m}(0)\!+\!{\cal W}_{\tilde m}( r_3)\!+\!{\cal W}_{\tilde m}( r_+)\!+\!{\cal W}_{\tilde m}( r_-)\right],
\end{align}
where $ r_\pm=(- r_3\pm\sqrt{3} r_8)/2$. The first term inside the bracket arises from the two color modes in the Cartan subspace whereas the last three terms are the contributions from the six transverse modes. It is easy to check that the previous expression is invariant under the symetries described above.

\begin{figure}[t!]  
\begin{center}
\hspace*{-.5cm}\epsfig{file=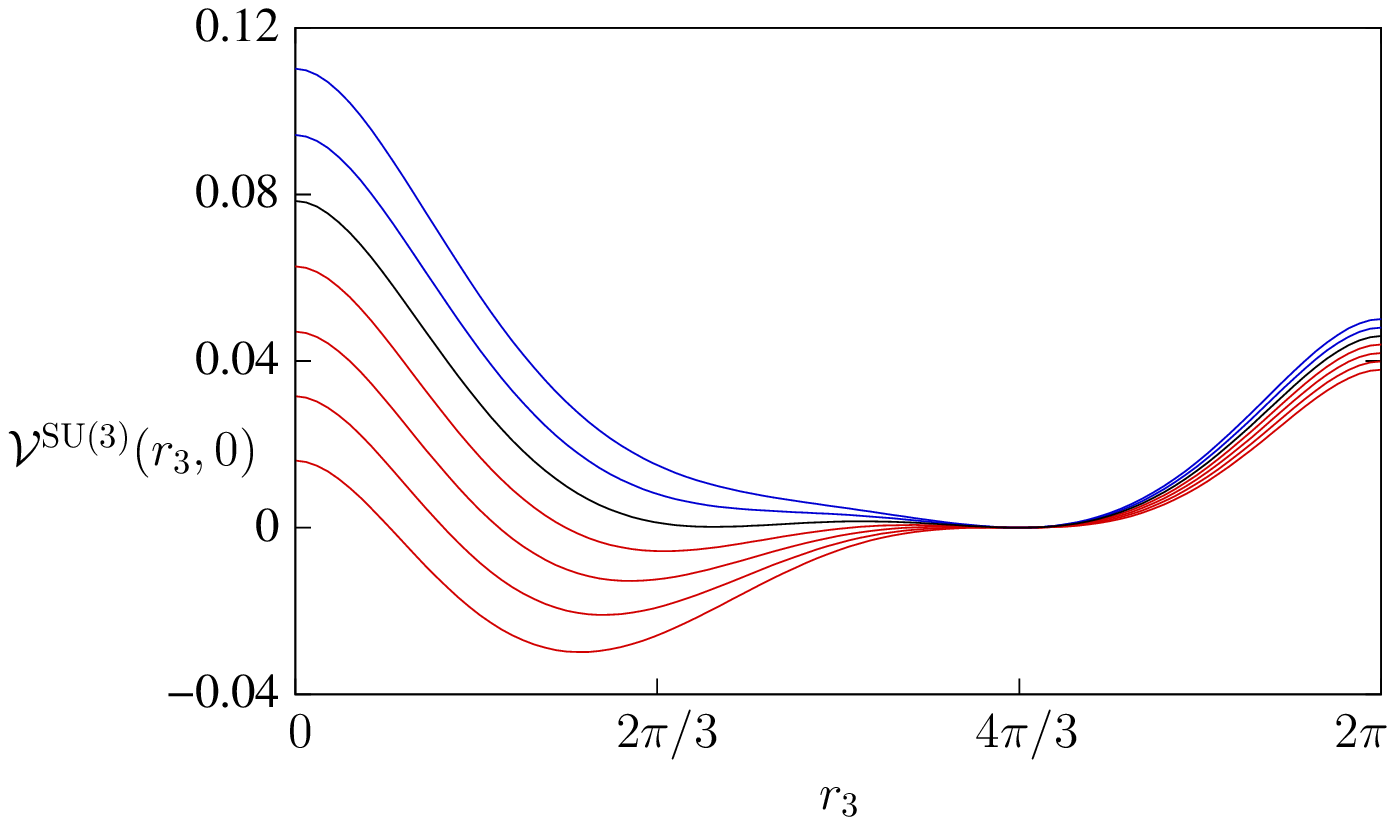,width=6.2cm}\\
\vspace{0.3cm}
\hspace*{-.5cm}\epsfig{file=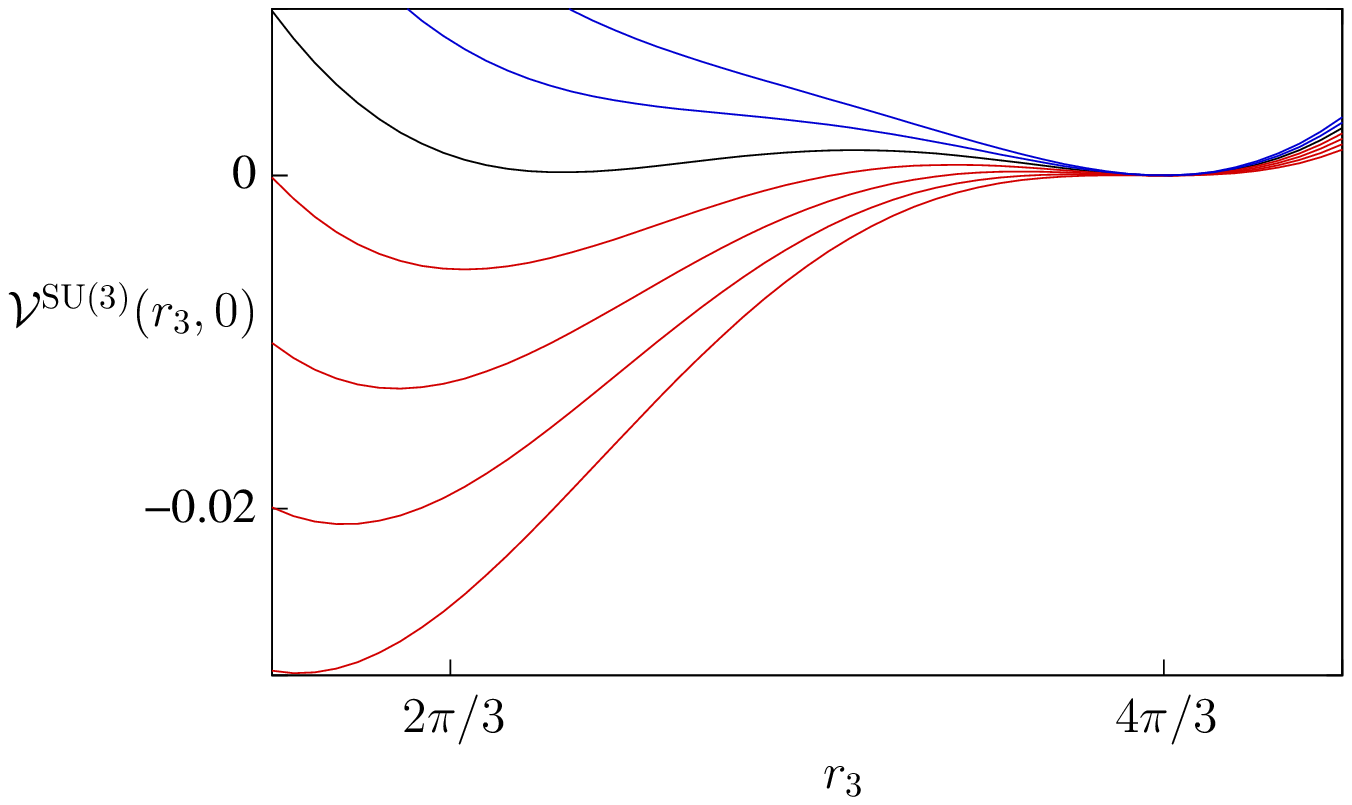,width=6.2cm}
 \caption{\label{fig:SU3pot} The SU($3$) background field potential in the $r_8=0$ direction in $d=4$, for decreasing values of $\tilde m$ (increasing temperatures) from top to bottom. The bottom figure is a close-up view around $T_{\rm c}$. (color online: $T=T_{\rm c}$ (black), $T<T_{\rm c}$ (blue), $T>T_{\rm c}$ (red))}
\end{center}
\end{figure}

At high temperatures, one recovers the Weiss potential~\cite{Weiss:1980rj}
\begin{align}
 {\cal V}^{\rm SU(3)}\approx(d-2)[F_0(0)\!+\!F_0(r_3)\!+\!F_0( r_+)\!+\!F_0( r_-)],
\end{align}
which has degenerate minima on the vertices of the basic equilateral triangle, corresponding to a deconfined phase. At low temperatures, we get, instead,
\beq
 {\cal V}^{\rm SU(3)}\approx-\left[F_0(0)\!+\!F_0(r_3)\!+\!F_0( r_+)\!+\!F_0( r_-)\right]
\eeq
and one easily checks that the absolute minimum lies at the center of the triangle, corresponding to a $Z_3$-symmetric, confining phase. A detailed analysis reveals that the transition is first order and that the absolute minimum lies on the $r_8=0$ axis (up to the aforementioned discrete rotations) for all temperatures. The potential \eqn{eq:potsu3} is plotted as a function of $r_3$ for $r_8=0$ in Fig.~\ref{fig:SU3pot} for various temperatures. We find the transition temperature $\tilde m_c\simeq 2.75$ in $d=4$. As before, we estimate the gluon mass parameter from fits of SU($3$) lattice data for the vacuum propagator in the Landau gauge, again its tree-level expression. In $d=4$, we find $m=510$~MeV which gives the estimate $T_c=185$~MeV. With the same words of caution as in the case $N=2$, we mention the results from lattice calculations \cite{Lucini:2012gg}, $T_c=270$~MeV, and from FRG studies \cite{Fister:2013bh}, $T_c=275$~MeV.

\begin{figure} 
\begin{center}
\epsfig{file=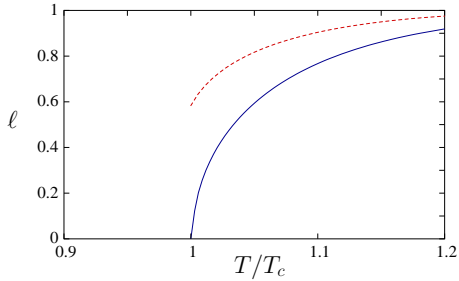,width=6cm}
 \caption{\label{fig:popol} The Polyakov loop as a function of the temperature normalized to the transition temperature for $N=2$ (solid line) and $N=3$ (dashed line). (color online) }
\end{center}
\end{figure}

Our results are summarized in \Fig{fig:popol}, which shows the temperature dependence of the Polyakov loop for both $N=2$ and $N=3$.

\section{The LDW gauge on the lattice, Gribov ambiguities and the gluon mass}

Let us now briefly discuss the possible relation between the gluon mass introduced in \eqn{eq_CF} and the Gribov ambiguities of the LDW gauge.
Writing the field $A=\bar A+a$, the gauge invariance of the Yang-Mills theory can be decomposed as
\beq 
\label{eq:sym1}
 \bar A_\mu\to \bar A_\mu^U\quad{\rm and}\quad
 a_\mu\to Ua_\mu U^{-1},
\eeq
or, equivalently, as 
\beq
\label{eq:sym2}
 \bar A_\mu\to \bar A_\mu\quad{\rm and}\quad a_\mu\to a_\mu^U=Ua_\mu U^{-1} +\frac{i}{g}U\bar D_\mu U^{-1},
\eeq
where $\bar D_\mu U=\partial_\mu U-ig[\bar A_\mu,U]$. The LDW condition \eqn{eq:LdW} is covariant under \eqn{eq:sym1} -- hence the symmetry \eqn{eq_ginv} of the gauge-fixed effective action -- but breaks the symmetry \eqn{eq:sym2}. 

The transformation $a\to a^U$ ressembles closely the usual gauge transformation with the replacement $\partial_\mu\to\bar D_\mu$. Observe that the LDW gauge condition and the complete gauge-fixed action \eqn{eq_gf} can be formally obtained from that in the Landau gauge by the same replacement. This can be used as a starting point for a nonperturbative formulation of the LDW gauge, as proposed in \cite{Cucchieri:2012ii}. Indeed, it is well-known that the Landau gauge condition can be obtained by extremizing the functional $\int_x\tr\{A_\mu^UA_\mu^U\}$ with respect to $U$ for a given $A$, which is the basis of efficient numerical implementations in lattice calculations. A simple generalization to the LDW gauge reads \cite{Cucchieri:2012ii}
\beq
\label{eq:minfunc}
 { F}_{\bar A}[a,U]=\int_x \tr \left\{a_\mu^U a_\mu^U\right\}.
\eeq
It is easy to check that the extrema of this functional with respect to $U$ for fixed $a$ and $\bar A$ satisfy the LDW condition \eqn{eq:LdW}. The authors of \cite{Cucchieri:2012ii} have demonstrated that background field gauges can be efficiently implemented in lattice calculations by means of similar methods as those routinely employed in the (minimal) Landau gauge. 

Continuum approaches face the usual Gribov problem, namely the fact that the functional \eqn{eq:minfunc} has many extrema (among which many minima). The gauge-fixed action \eqn{eq_gf} is based on the standard Fadeev-Popov construction which ignores Gribov copies and is, at best, a valid procedure in the high momentum regime. This is not a problem for lattice calculations since it is always possible to pick up one (local) minimum per gauge configuration, as done in the minimal Landau gauge. However, no such procedure can be formulated in the continuum in terms of a local and renormalizable action \cite{Gribov77}. 

Instead, the recent proposal of Refs. \cite{Serreau:2012cg,Serreau:2013ila} consists in taking a particular average over all copies of a given gauge orbit. It was shown in \cite{Serreau:2012cg} for the case of the Landau gauge that such a gauge-fixing procedure can be formulated in terms of a local, perturbatively renormalizable action in $d=4$.\footnote{This has been generalized in \cite{Serreau:2013ila} to a broader class of (nonlinear) covariant gauges.}  It was further shown that, for what concerns the calculation of gluon and ghost correlators, this gauge-fixed theory is perturbatively equivalent to the CFFT model mentioned in the introduction. In fact, the averaging procedure in \cite{Serreau:2012cg} involves the Landau gauge extremization functional mentioned above, which eventually leads to the effective gluon mass term.\footnote{More precisely, the effective bare mass has to be sent to zero together with the (continuum) limit of vanishing bare coupling. This can be done by keeping the renormalized mass finite.}

It is straightforward to generalize the proposal of \cite{Serreau:2012cg} to the LDW gauge by using \eqn{eq:minfunc} in place of the Landau gauge extremization functional. Eventually, this amounts to the replacement $\partial_\mu\to\bar D_\mu$ described previously. In particular, it can be shown that the main results hold: the gauge-fixed theory can be formulated in terms of a local action, perturbatively renormalizable in $d=4$, and it effectively reduces to the massive action \eqn{eq_CF} for what concerns the calculation of gluon and ghost correlators and, consequently, of gauge invariant observables.

\section{Conclusions}

We have shown that the phase structure of SU($N$) Yang-Mills theories can be described in a perturbative set-up at the first non-trivial, one-loop order. This is a remarkable result. Our approach is based on a simple massive modification of the usual Fadeev-Popov action in the LDW gauge, where the mass term can be related to the issue of properly taking into account Gribov ambiguities. Using values of the mass parameter from fits of lattice data for gluon correlators at vanishing temperature, we obtain rather satisfying results for the transition temperatures for $N=2$ and $N=3$ in view of the simplicity of the calculation. We mention that our one-loop result compares qualitatively well with those of \cite{Braun:2007bx,Braun:2010cy} using the FRG approach. This demonstrates that the main nonperturbative physics associated with the deconfinement transition -- and thus with the confinement of static quarks -- is well captured by a simple gluon mass term and that a perturbative treatment around the massive theory is adequate. 

This has various welcome consequences: First, the relevant physics can be efficiently described by simple (perturbative) calculations. For instance, an interesting extension of the present work is to include quarks at finite baryonic chemical potential at one loop order. This will be presented elsewhere. Second, the approximation can be systematically improved by going to higher orders. For instance, it is a hard but doable calculation to improve on the results presented here by going to two-loop order. Preliminary results show that the critical temperature in the SU($2$) case comes closer to its actual value and that the thermodynamic pressure is better behaved as a function of the temperature. This, also, will be the object of a separate publication. 

We mention that we have computed the pressure \eqn{eq:pressure} as a function of temperature. The one-loop result has several unwanted features such as the fact that it behaves as $T^4$ (in $d=4$) at low temperatures and that it briefly decreases with temperature around the phase transition. Our preliminary two-loop results seem to resolve this second problem, see \cite{Reinosa:2014zta} for the SU($2$) case. Contrarily to the Polyakov loop, whose behavior is governed by symmetry, the pressure is sensitive to the full Yang-Mills dynamics, which probably explains why it is more difficult to compute accurately.

Finally, we emphasize that the ability of the present approach to efficiently capture the essential aspects of the phase transition is deeply related to the fact that the center symmetry is manifest at each step of the calculation. In contrast, the $Z_N$ symmetry is not explicit in the Landau gauge. This might be related to the poor convergence of lattice data for the chromoelectric gluon propagator in the Landau gauge near the transition temperature \cite{Mendes:2014gva}. It woud be interesting to study the possibility of allowing for a nontrivial background temporal gluon field in lattice calculations of Yang-Mills correlators at finite temperature by implementing a discrete version of the LDW gauge, e.g., along the lines of \cite{Cucchieri:2012ii}.

\section*{Acknowledgements}

We would like to thank L. Fister, T. Mendes and J.M. Pawlowski for interesting discussions on topics related to this work. We also acknowledge support from the ECOS cooperation program and finantial support from the PEDECIBA program.

\appendix

\section{Symmetries}\label{sym}

For completeness we briefly recall the basic constraints from the $Z_N$ symmetry and the charge conjugation invariance.
We consider gauge transformations of the form $U(\tau,{\bf x})=\exp\left\{i\tau\varphi\right/\beta\}$, with  $\varphi=\varphi_a t^a_C$ an element of the Cartan algebra. For appropriate values of $\varphi$, these preserve both the $\beta$-periodicity in imaginary time and the spatial homogeneity of the background field and leave it in the temporal direction and in the Cartan algebra.  
Under such a gauge transformation, the parameters $r_a=\beta g \bar A_0^a$ get shifted as $r_a\to r_a+\varphi_a$. Depending on the allowed values of the angles $\varphi_a$, the background field gauge invariance thus leads to periodicity properties of the potential \eqn{eq:poteff}. Below we examine the constraints from $Z_N$ transformations, which correspond to gauge transformations with twisted boundary conditions, $U^\dagger(\beta)U(0)=e^{2i\pi/N}\mathds{1}$.

We also exploit the invariance of the (gauge-fixed) theory under the charge-conjugation transformation $(A_\mu^at^a)\to -(A_\mu^at^a)^*$ which leads to a parity symmetry $r_a\to-r_a$ of the potential \eqn{eq:poteff} in the Cartan algebra.

In the case $N=2$, the Cartan algebra has a single generator $t_3={1\over2}{\rm diag}(1,-1)$. One easily checks that periodic gauge transformations, such that $U^\dagger(\beta)U(0)=\mathds{1}$ correspond to $\varphi_3=4\pi$ whereas twisted ($Z_2$) transformations lead to a shorter $2\pi$-periodicity. Together with charge conjugation invariance, this leads to the symmetry property \eqn{eq:symSU2}.

In the case $N=3$, the two generators of the Cartan algebra can be chosen as $t_3={1\over2}{\rm diag}(1,-1,0)$ and $t_8={1\over2\sqrt{3}}{\rm diag}(1,1,-2)$. The presence of the vanishing eigenvalue of $t_3$ prevents the possibility of a twisted gauge transformation. Only standard periodic gauge transformation are possible in the $r_3$-direction, which leads to a $4\pi$-periodicity. In the $r_8$-direction, twisted $Z_3$ transformations are possible and lead to a shorter $4\pi/\!\sqrt{3}$-periodicity. The other symmetries of the potential follow from charge conjugation invariance and the global color symmetry. They imply in particular that $r_3=0$, $r_8=0$, $r_8=\pm r_3/\!\sqrt{3}$ and  $r_8=\pm r_3\sqrt{3}$ are symmetry axis.

\end{document}